\documentclass{article}
\usepackage{spconf,amsmath,graphicx}
\usepackage{amssymb}
\usepackage{algorithm,algorithmic}
\usepackage{xcolor}
\usepackage{harpoon}

\mathchardef\mhyphen="2D
\def\cpar{\hss\egroup\line\bgroup\hss}

\title{Ultra-Sparse View Reconstruction for Flash X-Ray Imaging using Consensus Equilibrium}
\name{
Maliha Hossain$^1$, 
Shane C. Paulson$^2$, 
Hangjie Liao$^3$, 
Wienong W. Chen$^2$,
Charles A. Bouman$^1$ \thanks{This work was supported by NSF grant number CCF-1763896.}
}
\address{
$^1$School of Electrical and Computer Engineering, Purdue University, West Lafayette, IN, USA, \\ 
$^2$School of Aeronautics and Astronautics, Purdue University, West Lafayette, IN, USA, \\
$^3$Lam Research, 4400 Cushion Pkwy, Fremont, CA 94538, USA
}
\begin{document}
%
\maketitle 

\begin{abstract}
A growing number of applications require the reconstruction of 3D objects from a very small number of views.
In this research, we consider the problem of reconstructing a 3D object
from only 4 Flash X-ray CT views taken during the impact of a Kolsky bar.
For such ultra-sparse view datasets, even model-based iterative reconstruction (MBIR) methods
produce poor quality results.

In this paper, we present a framework based on a generalization of Plug-and-Play, known as Multi-Agent Consensus Equilibrium (MACE), 
for incorporating complex and nonlinear prior information into ultra-sparse CT reconstruction.
The MACE method allows any number of agents to simultaneously enforce their own prior constraints on the solution.
We apply our method on simulated and real data and demonstrate that MACE reduces artifacts, improves reconstructed image quality, 
and uncovers image features which were otherwise indiscernible.

\end{abstract}
\begin{keywords}
Sparse View Tomography, 
In-Situ Tomography,
Consensus Equilibrium, 
Inverse Problems
\end{keywords}
\section{Introduction}
\label{sec:intro}

High speed volume visualization of dynamic processes is 
an emerging application of computed tomography (CT). 
In particular, Flash X-ray has been used to study 
explosive and ballistic events \cite{tringe2019dynamic, moser2019investigation},
material deformation \cite{moser2014situ},
and production line quality control \cite{bauza2018realization}, 
where space and time constraints necessitate sparse view sampling. 
Even in the absence of such constraints, sparse view CT presents benefits 
in the form of reduced computation and memory requirements, 
as well as lower radiation exposure in medical CT.

Conventional CT reconstruction, with a few thousand views, is often performed using analytical reconstruction methods such as Filtered Back Projection (FBP).
However, these analytical methods typically generate severe artifacts for sparse view CT problems in which the number of views is subsampled by a factor of 2 to 10.
For sparse-view CT, methods such as model based iterative reconstruction (MBIR) must be used to achieve high quality results \cite{bouman1993generalized}.

When the number of views is less than 10,
we enter the domain of ultra sparse CT, and it becomes very challenging to reconstruct high quality CT images.
In this case, even established methods such as MBIR tend to generate severe artifacts with reduced contrast and resolution.
Recently, Henzler et. al \cite{henzler2018single} and Shen et. al \cite{shen2019patient} have independently developed methods
for inferring 3D reconstructions from a single medical view using deep learning methods.
However, Flash X-ray CT events are typically destructive and stochastic in nature.
This makes it difficult to collect the data that would be required to train the machine learning methods.
In other recent work, Moser et. al \cite{moser2014situ, moser2019investigation} used the algebraic reconstruction technique (ART)
to reconstruct 3D volumes from as few as 6 views.
However, ART does not allow for the explicit modeling of prior distributions or even the sensor forward model.

Traditional MBIR formulates the reconstruction problem as the solution to a
maximum a posteriori (MAP) or regularized maximum likelihood optimization problem \cite{bouman1993generalized}.
More recently, Plug and Play (PnP) methods \cite{venkatakrishnan2013plug}, originally based on ADMM optimization algorithms \cite{sidky2012convex, ramani2011splitting, zheng2017sparse},
have expanded MBIR by allowing the prior distribution to be replaced with advanced denoising operations \cite{balke2018separable,sreehari2015advanced}.
Notably, BM3D \cite{dabov2006image} and its higher dimensional variants are among the best known denoisers that do not require training,
and have been found to work particularly well with PnP.

Most recently, Multi-Agent Consensus Equilibrium, or MACE \cite{buzzard2018plug} is a generalization of PnP
that is based on the solution of a set of equilibrium equations.
The MACE equilibrium has the interpretation of being the consensus solution reached by multiple agents
each operating in a way that enforces its own preference or constraint.
A discussion of practical implementation considerations 
for fast execution is presented in \cite{sridhar2019distributed}.
Multi-Slice Fusion \cite{majee20194d, majee2020multi} is a variant of MACE,
where multiple lower dimensional denoisers are fused to form a higher dimensional prior model.

In this paper, we introduce a new algorithm, multi-slice fusion with rotational invariance (MSF-RI) for the reconstruction of ultra-sparse tomographic data;
and we apply the MSF-RI algorithm to the reconstruction of 4 view CT data from a flash X-ray in-situ imaging system used in a Kolsky bar impact experiment  \cite{liao2018flash}.
Our algorithm works by integrating the constraints of four distinct agents using the MACE framework.
The first three agents are based on 2D regularization using BM3D, and the fourth agent enforces partial rotational invariance using a rotational smoothing operator.
Our experiments compare MSF-RI to FBP, traditional MBIR, and PnP using a BM4D prior on both synthetic and real data sets.
Our proposed MSF-RI method achieves better RMSE and SSIM values than previous methods
and reveals structures in our volumetric reconstructions
that were previously indistinguishable from streak artifacts.


\section{Problem Formulation}
\label{sec:formulation}

A CT scan is a collection of X-ray measurements of an object taken at different angles.
Reconstructing the object from the scan is an inverse problem with a forward model of the form
$$
y = Ax + noise
$$
where $y\in \mathbb{R}^{M}$ is the measurement vector,
$A\in \mathbb{R}^{M\times N}$ is the scanner system matrix,
and $x\in \mathbb{R}^{N}$ is the latent image vector we wish to recover.
The measurements contain additive Gaussian noise distributed as $N(0,\alpha \Lambda^{-1})$.
The MAP cost function then has the form
\begin{equation}
\hat{x}_{MAP} = \arg\min_{x}\{f(x) + h(x)\}
\end{equation}
with a forward model term, $f(x) =-\log p(y|x)$, to enforce data fidelity,
and a prior model term, $h(x)=-\log p(x)$ to impose regularity.
In this case, the forward model term can be written within a constant as
\begin{equation}
    f(x) = \frac{1}{2\alpha} ||y - Ax||_{\Lambda}^{2} \ .
\end{equation}
In practice, it is often difficult to express very complex prior distributions as a single tractable functions $h(x)$.
Therefore, the remainder of this section outlines how we will replace this prior
with one tailored for our application using MACE.

\subsection{MACE framework}
\label{ssec:maceframwork}

Consider the MAP cost function with $K$ regularizers 
\begin{equation}
    \hat x_{MAP} = \arg \min_{x}\left\{ f(x) + \sum_{k=1}^{K} \beta_k \, h_k(x) \right\}
\label{eq:MAP-equation}
\end{equation}
where $\beta_k\geq 0$ controls the amount of regularization applied by $h_k$.
The MACE algorithm splits this MAP estimation problem into pieces, with each piece enforced using an agent.
For the exact MAP estimation problem, the forward model agent is the proximal map given by
\begin{equation}
    F(w) = \arg \min_{z\in\mathbb R^N} \left\{f(z) + \frac{1}{2\sigma^2} ||z-w||^2 \right\}
\label{eq:inv_operator}
\end{equation}
and the prior model agents are given by
\begin{equation}
    H_k(w) = \arg \min_{z\in\mathbb R^N} \left\{h_k(z) + \frac{1}{2\sigma^2} ||z-w||^2 \right\}
\label{eq:H-operators}
\end{equation}
where $\sigma > 0$ controls the convergence speed of the algorithm.

Like PnP, MACE allows us to replace the proximal maps $H_1 , \dots, H_K$ with more general denoising agents that typically are not proximal maps.
Moreover, MACE is a generalization of PnP \cite{buzzard2018plug} because it allows for the case of multiple agents, which we will use in this work.
 Additionally, MACE specifies the solution to the problem using an equilibrium condition, rather than as a solution to an optimization problem.
This is important since, in general, the PnP or MACE solution is no longer the solution to a MAP style optimization problem when the agents are not proximal maps.

In order to write the MACE equilibrium condition, define the following notation.
Let $\mathbf w = [w_0 , w_1, \dots, w_K]\in \mathbb R^{N\times (K+1)}$  be the concatenation of $K+1$ states 
where each state will serve as the input to its corresponding agent.
Next we define the agent operator, 
$$
L(\mathbf w)=[ F(w_0) , H_1(w_1), \dots, H_K(w_K) ] \ ,
$$
to denote the parallel application of the $K+1$ agents on their respective state inputs. 
We also define the averaging operator $G(\mathbf w)=[\bar{w}, \dots, \bar{w}]$,
where $\bar{w} = \sum_{k=0}^K \mu_k w_k$.
The vector $\mu\in \mathbb R^{K+1}$ controls  the influence each agent exerts on the equilibrium solution and is given by
\begin{equation}
    \mu = \frac{1}{1+\sum_{k=1}^K\beta_k}\left[ 1 , \beta_1, \dots, \beta_K \right]
\end{equation}

\vspace{-0.2cm}
\begin{algorithm}[b]
 \caption{MACE algorithm}
 \label{table:MACE-Algorithm}
 \begin{algorithmic}[1]
 \renewcommand{\algorithmicrequire}{\textbf{Input:}}
 \renewcommand{\algorithmicensure}{\textbf{Output:}}
 \REQUIRE Initial Reconstruction: $x^{(0)} \in \mathbb R^N$
 \ENSURE  Final Reconstruction: $x^{*}$
 \\ \STATE $\mathbf w \gets [x^{(0)},\dots, x^{(0)}] $
  \WHILE {not converged}
  \STATE $\mathbf x \gets  L ( \mathbf w )$
  \STATE $\mathbf z \gets  G(2\mathbf x - \mathbf w)$
  \STATE $\mathbf w \gets \mathbf w + 2\rho(\mathbf z - \mathbf x)$
  \ENDWHILE
 \RETURN $x^{*} \gets \sum_{k=1}^{K+1}\mu_k x_k$ 
 \end{algorithmic} 
\end{algorithm}

Using this framework, the solution to our problem is given by $\hat{x} = \sum_{k=0}^K \mu_k w_k^*$,
where $w^*$ is the solution to the consensus equilibrium (CE) equation given by
\begin{equation}
L(\mathbf w^* ) = G(\mathbf w^* ) \ .
\label{eq:CE-condition}
\end{equation}
We note that when the agents are the proximal maps of (\ref{eq:inv_operator}) and (\ref{eq:H-operators}),
then $\hat{x}=\hat{x}_{MAP}$ is exactly the MAP estimate of (\ref{eq:MAP-equation}).

In~\cite{buzzard2018plug}, it is shown that the CE equations can be solved by using the Douglas-Rachford algorithm 
to solve for the fixed point of the operator $T= (2G-I)(2F-I)$.
Algorithm~\ref{table:MACE-Algorithm} shows the general method for doing this.
When $K=1$, and $\rho = 0.5$, this algorithm corresponds exactly to PnP implemented with the consensus ADMM algorithm \cite{buzzard2018plug},
and a sufficient condition for convergence is that $T$ is non-expansive.

\subsection{MACE Agents}
\label{ssec:maceagents}

Our reconstruction algorithm uses five agents.
The first agent, $F$, enforces data fidelity and is defined in (\ref{eq:inv_operator}) 
and solved using Iterative Coordinate Descent.

The agents $H_1$ through $H_4$ are chosen to enforce spatial regularity and together act as a prior model for reconstruction.
More specifically, agents $H_1$, $H_2$, and $H_3$ are 2D BM3D denoisers applied to the
$(\overrightharp{x},\overrightharp{y})$, 
$(\overrightharp{y},\overrightharp{z})$, 
and $(\overrightharp{z},\overrightharp{x})$ planes respectively.
The $H_1$ agent independently applies the BM3D denoiser to each 
$(\overrightharp{x},\overrightharp{y})$ slice of the 3D volume.
$H_2$ and $H_3$ perform similar functions along slices in
$(\overrightharp{y},\overrightharp{z})$, and 
$(\overrightharp{z},\overrightharp{x})$.
We refer to this integration of three 2D denoising algorithms as multi-slice fusion (MSF) as described in more detail in~\cite{majee20194d}.

Agent $H_4$ is designed to enforce weak rotational invariance in the reconstruction.
By weak rotational invariance we mean that the reconstructed object should be approximately invariant to small rotations around its axis of symmetry.
This is physically reasonable since both the sample under test and the experimental forces are approximately rotationally invariant.
Nonetheless, the actual sample will exhibit asymmetric cracks of importance, so $H_4$ acts to weakly regularize the solution, rather than to enforce strict rotational symmetry.

The weak rotational invariance agent is defined by
\begin{align*}
    H_4(x) = \sum_{n = -J}^{J}\gamma_n \,R_{ n (\theta /J) }(x)
\end{align*}
where $R_{\phi}(x)$ rotates the 3D volume, $x$, around its axis of rotational symmetry by an angle of $\phi$.
Intuitively, $H_4$ performs a weighted average of rotations over an angle of $2\theta$ degrees, with weights, $\gamma_j$, drawn from a scaled Hamming window.

\section{RESULTS}
\label{sec:results}

In this section, we present both simulated and real data reconstruction results 
for a Flash X-ray CT scanner that acquires four views oriented at $18^{\circ}$, $162^{\circ}$, $234^{\circ}$, and $306^{\circ}$ \cite{liao2018flash}
using the following five reconstruction algorithms:

\begin{table}[!ht]
\begin{tabular}{lll}
& {\em Algorithm} & {\em Description} \\
& FBP & Filtered Back Projection \\
& qGGMRF & MBIR reconstruction is a qGGMRF prior  \\
& PnP-BM4D & PnP prior using the 3D denoiser BM4D \\
& MSF & MACE with agents $H_1$, $H_2$, and $H_3$ \\
& MSF-RI & MACE with agents $H_1$, $H_2$, $H_3$, and $H_4$ \\
\end{tabular}
\end{table}
\smallskip

\noindent
For the simulated cylinder, results and parameters are 
chosen to minimize RMSE scores achieved using the different priors. 
For the experimental case, the most visually pleasing results are reported. 
In all instances of MACE, we set $\rho = 0.4$, and $\theta = 8^{\circ}$ for $H_4$, where applicable.
The block matching modules use the implementation in \cite{bm3dcode} and \cite{bm4dcode} with default parameters. 
The PnP and MACE algorithms are all initialized using a qGGMRF reconstruction. 

\begin{figure}[htb]
\begin{minipage}[b]{1.0\linewidth}
  \centering
  \centerline{\includegraphics[width=5.0cm]{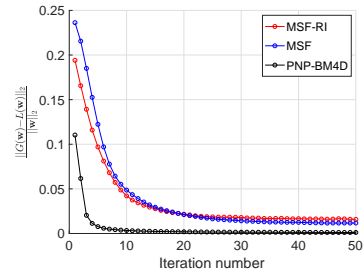}}
\end{minipage}
\vspace{-0.6cm}
\caption{Convergence rates for phantom reconstruction with PnP-BM4D, MSF, and MSF-RI}
\label{fig:sim_convergence_new}
\end{figure}

\begin{figure}[htb]
\begin{minipage}[b]{1.0\linewidth}
  \centering
  \centerline{\includegraphics[width=5.0cm]{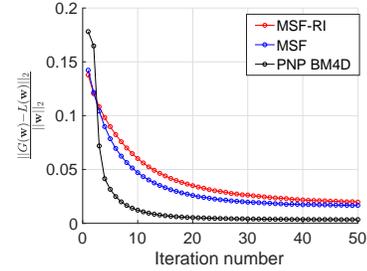}}
\end{minipage}
\vspace{-0.6cm}
\caption{Convergence rates for reconstruction from of real dataset with PnP-BM4D, MSF, and MSF-RI}
\label{fig:real_convergence_new}
\end{figure}

Convergence plots are shown for both datasets 
in Figures~\ref{fig:sim_convergence_new} and~\ref{fig:real_convergence_new}.
Distance to convergence is quantified as the RMSE between $G(\mathbf w)$ and $L(\mathbf w)$ normalized over $\mathbf w$.
This value reflects how closely equation~(\ref{eq:CE-condition}) is satisfied. 
We see that all PnP and MACE algorithms 
are able to converge to within less than $5\%$ of the final solution.


\begin{figure}[t]
\centering
\begin{minipage}[b]{.32\linewidth}
  \centering
  \centerline{\includegraphics[height=2.5cm]{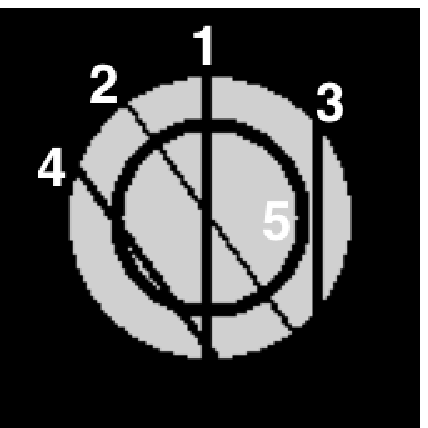}}
  \vspace{-0.1cm}
  \centerline{(a) Ground Truth}\medskip 
\end{minipage}
\hfill
\begin{minipage}[b]{0.32\linewidth}
  \centering
  \centerline{\includegraphics[height=2.5cm]{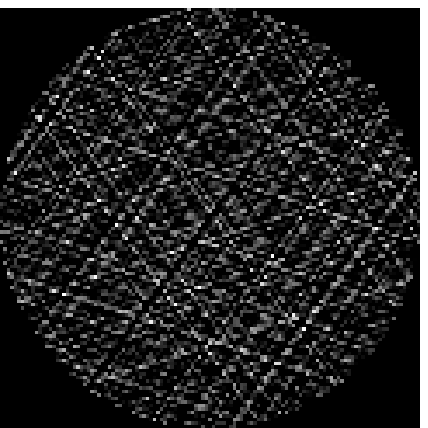}}
  \vspace{-0.1cm}
  \centerline{(b) FBP}\medskip 
\end{minipage}
\hfill
\begin{minipage}[b]{.32\linewidth}
  \centering
  \centerline{\includegraphics[height=2.5cm]{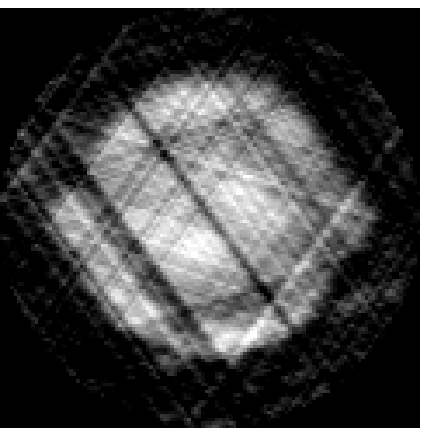}}
  \vspace{-0.1cm}
  \centerline{(c) qGGMRF}\medskip 
\end{minipage}
\hfill
\begin{minipage}[b]{0.32\linewidth}
  \centering
  \centerline{\includegraphics[height=2.5cm]{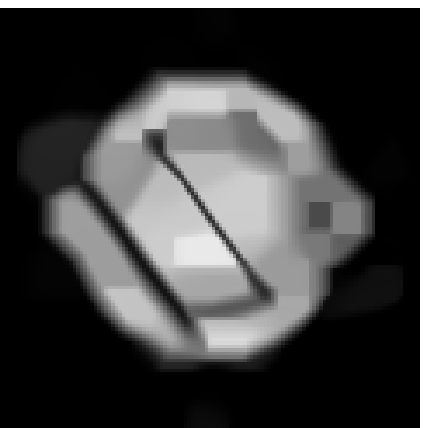}}
  \vspace{-0.1cm}
  \centerline{(d) PnP-BM4D}\medskip 
\end{minipage}
\hfill
\begin{minipage}[b]{.32\linewidth}
  \centering
  \centerline{\includegraphics[height=2.5cm]{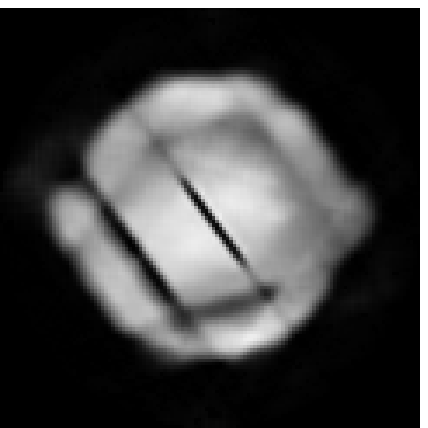}}
  \vspace{-0.1cm}
  \centerline{(e) MSF}\medskip 
\end{minipage}
\hfill
\begin{minipage}[b]{0.32\linewidth}
  \centering
  \centerline{\includegraphics[height=2.5cm]{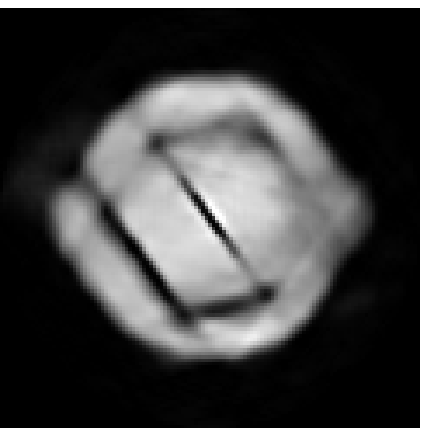}}
  \vspace{-0.1cm}
  \centerline{(f) MSF-RI}\medskip 
\end{minipage}
\hfill
\vspace{-0.4cm}
\caption{
Simulated results.  (a) sample cross section of ground truth for comparison 
with CT reconstructions (b) through (f) with various methods. 
(d), (e), and (f) are initialized with the qGGMRF result in (c).
}
\label{fig:sim_results}
\end{figure}
\vspace{-0.4cm}

\vspace{0.1cm}

\begin{figure*}
\begin{minipage}[b]{.17\linewidth}
  \centering
  \centerline{\includegraphics[height=2.5cm]{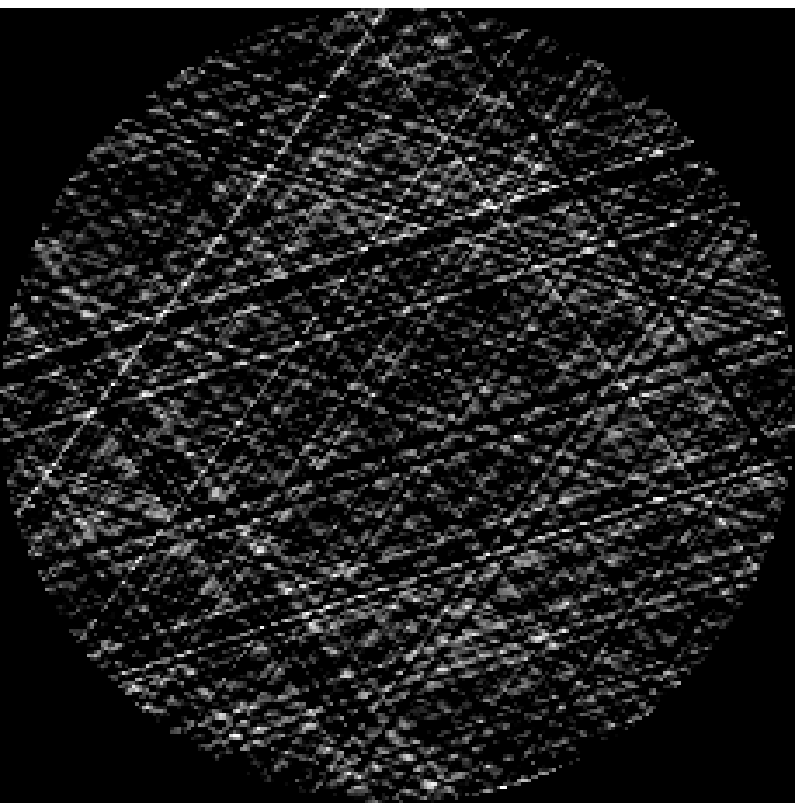}}
  \vspace{-0.1cm}
  \centerline{(a) FBP}\medskip
\end{minipage}
\hfill
\begin{minipage}[b]{.17\linewidth}
  \centering
  \centerline{\includegraphics[height=2.5cm]{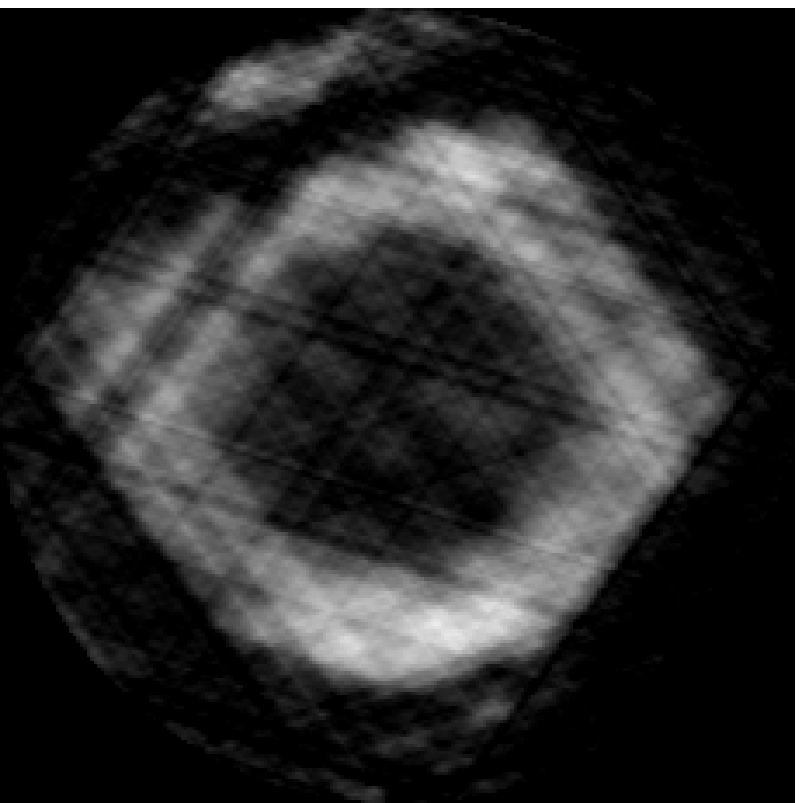}}
  \vspace{-0.1cm}
  \centerline{(b) qGGMRF}\medskip
\end{minipage}
\hfill
\begin{minipage}[b]{.17\linewidth}
  \centering
  \centerline{\includegraphics[height=2.5cm]{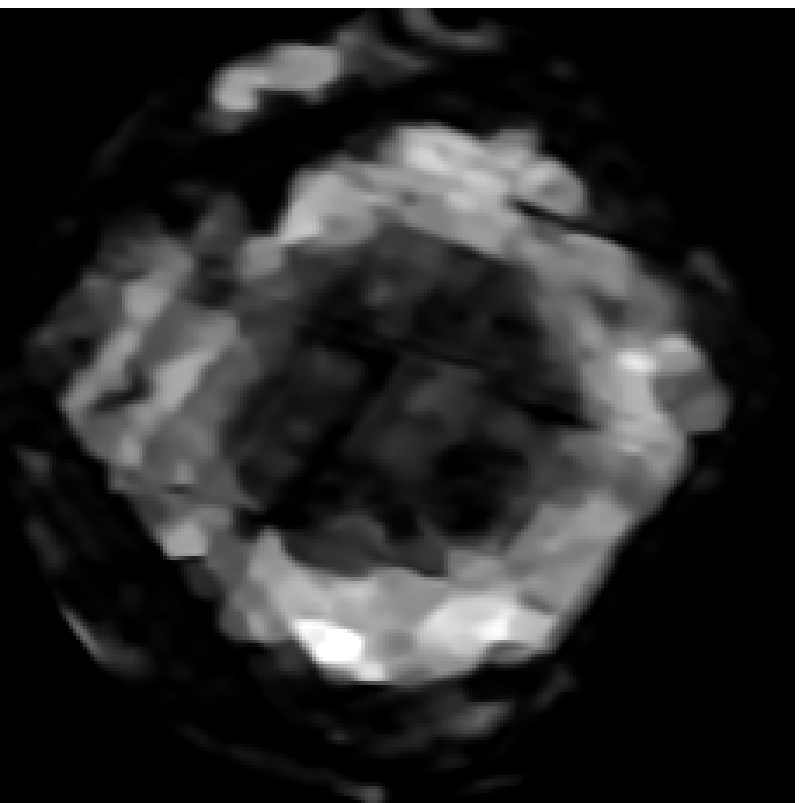}}
  \vspace{-0.1cm}
  \centerline{(c) PnP-BM4D}\medskip  
\end{minipage}
\hfill
\begin{minipage}[b]{0.17\linewidth}
  \centering
  \centerline{\includegraphics[height=2.5cm]{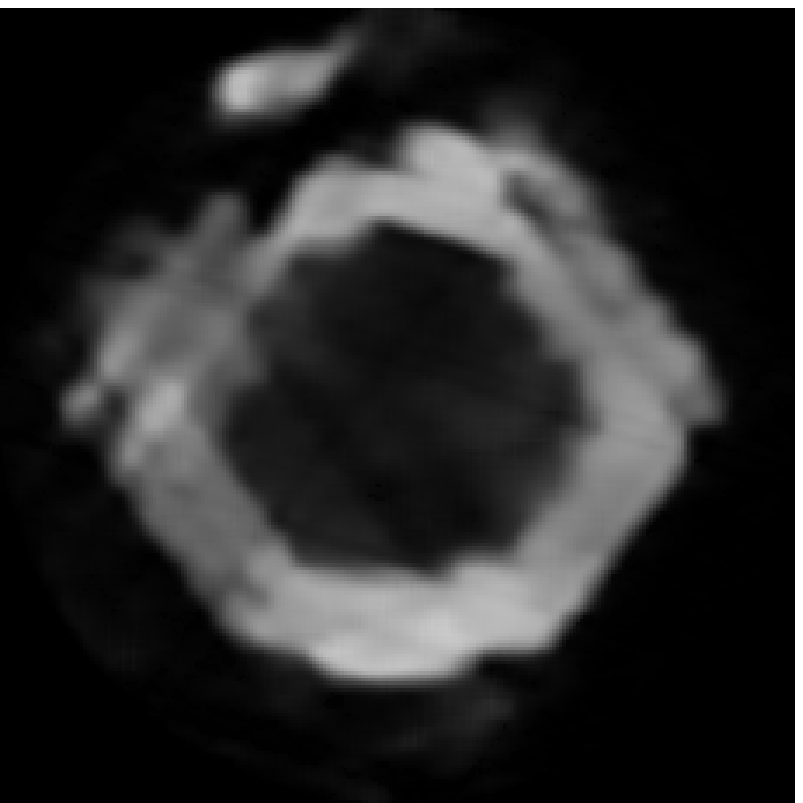}}
  \vspace{-0.1cm}
  \centerline{(d) MSF}\medskip
\end{minipage}
\hfill
\begin{minipage}[b]{.17\linewidth}
  \centering
  \centerline{\includegraphics[height=2.5cm]{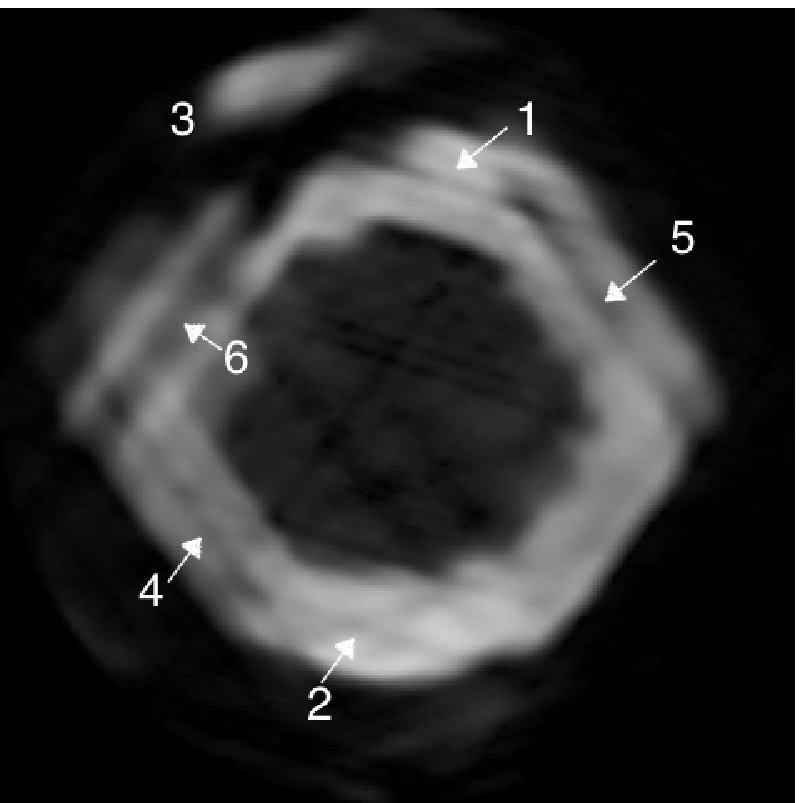}}
  \vspace{-0.1cm}
  \centerline{(e) MSF-RI}\medskip  
\end{minipage}
\hfill
\vspace{-0.4cm}
\caption{Reconstructions of slice 21 from experimental dataset. Features of interest are numbered in the result of the proposed method for cross referencing with measurement data.}
\label{fig:real_results}
\end{figure*}

\vspace{-0.3cm}
\subsection{Simulated dataset}
Figure~\ref{fig:sim_results}(a)
shows the simulated cylinder phantom
with transverse, and radial cracks, typically present in concrete specimens
that have been loaded in a Kolsky bar experiment \cite{paulson_dymat2020}.
The volume is $121 \times 121$ voxels by 100 slices. 
Cracks numbered 2 and 4 lie along the view angle at $234^{\circ}$, 
while 1 and 3 are oriented at $90^{\circ}$.
The concentric crack, labelled 5, has been added to further test the model's capabilities. 
The 3D phantom was forward projected with the parallel beam $A$ matrix
and corrupted with AWGN at $20\%$ of the signal strength. 

\begin{table}[htb]
\caption{Metrics for simulated dataset reconstruction.}
\vspace{0.3cm}
\label{tab:rmse_ssim_score}
\centering
\begin{tabular}{|c|c|c|}
\hline
Method   & NRMSE & SSIM              \\ \hline
FBP      & 0.8678             & 0.275052          \\ \hline
qGGMRF   & 0.2389             & 0.625456          \\ \hline
PNP-BM4D & 0.2044             & 0.778083          \\ \hline
MSF      & 0.2006             & 0.778822          \\ \hline
MSF-RI   & \textbf{0.1909}    & \textbf{0.789262} \\ \hline
\end{tabular}
\end{table}


Figures~\ref{fig:sim_results}(b) through (g) 
show CT reconstructions of the sample made using the five algorithms,
and Table~\ref{tab:rmse_ssim_score} lists the associated normalized root mean squared error (NRMSE) and SSIM values reported for the entire volume.

Notice that FBP fails, and even traditional MBIR performs poorly on this very underdetermined dataset.
PnP-BM4D shows improved results, but generates blocky artifacts.
This is likely due to the lack of similar patches 
in three dimensions as opposed to two dimensions
in such a small volume.
Consequently, the multi-slice fusion used in MSF is more effective than the BM4D prior,
and the result shows more detail.
The best result, both visually, and numerically is the MACE-RI algorithm, which reveals the most detail,
including the concentric crack~5, and the transverse cracks~2 and~4.
Notice that all the methods are effectively blind to crack~1, and crack~3 appears as a blur.

\begin{figure}[htb]
\begin{minipage}[b]{.48\linewidth}
  \centering
  \centerline{\includegraphics[width=3.7cm]{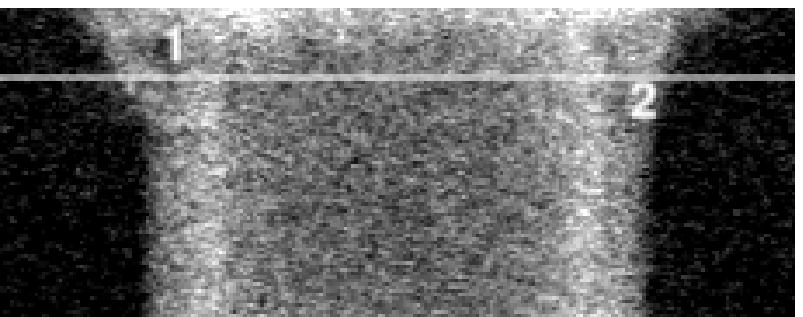}}
  \vspace{-0.1cm}
  \centerline{(a) View at $18^{\circ}$}\medskip
\end{minipage}
\hfill
\begin{minipage}[b]{0.48\linewidth}
  \centering
  \centerline{\includegraphics[width=3.7cm]{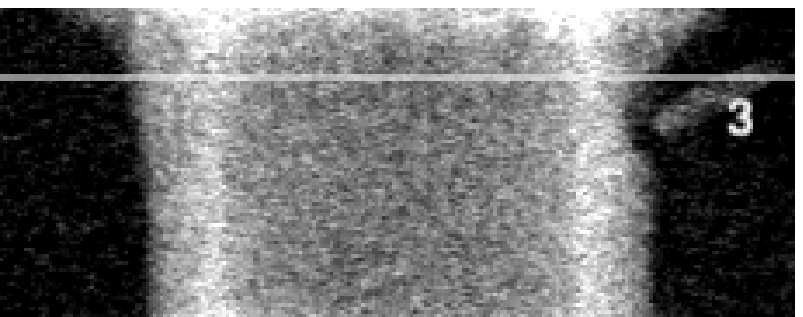}}
  \vspace{-0.1cm}
  \centerline{(b) View at $162^{\circ}$}\medskip
\end{minipage}
\begin{minipage}[b]{.48\linewidth}
  \centering
  \centerline{\includegraphics[width=3.7cm]{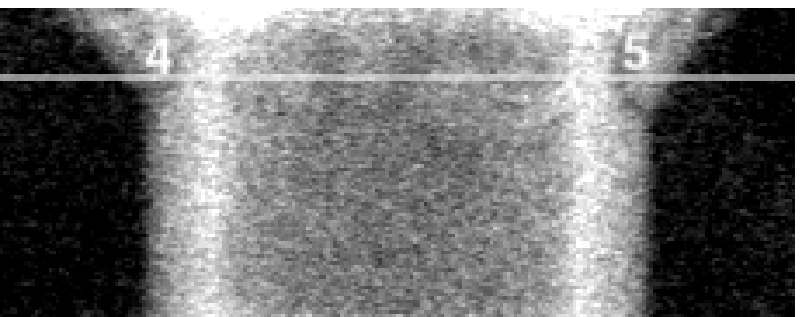}}
  \vspace{-0.1cm}
  \centerline{(c) View at $234^{\circ}$}\medskip
\end{minipage}
\hfill
\begin{minipage}[b]{0.48\linewidth}
  \centering
  \centerline{\includegraphics[width=3.7cm]{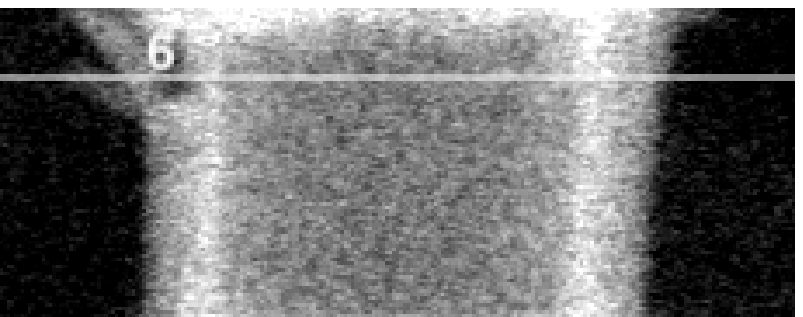}}
  \vspace{-0.1cm}
  \centerline{(d) View at $306^{\circ}$}\medskip
\end{minipage}
\vspace{-0.4cm}
\caption{Experimental view data, where the location of slice 21 is marked with a white line and features of interest are numbered for comparison in reconstructions.}
\label{fig:real_data}
\end{figure}

\vspace{-0.3cm}


\subsection{Experimental dataset}
\label{ssec:experimental}
The real sandstone ring dataset is imaged 
at the moment of impact in the Kolsky bar experiment.
The measurement data consists of four views of 478 detector channels per view and 198 slices.
After a $2\times 2$ binning factor, the reconstructed volume is $229\times 229$ by 89 slices with voxels of width 0.097 mm.
The reconstructions in Figure~\ref{fig:real_results} 
each depict slice 21 in the $(\overrightharp{x}, \overrightharp{y})$ plane, 
which is also marked in the preprocessed view data in
Figure~\ref{fig:real_data} by a white line.

Notice that the FBP and qGGMRF results again show severe streaking artifacts
that appear along the view angles 
due to the presence of noise in the projection data.
Once again, PnP-BM4D produces blocky regions and cannot resolve the finer cracks. 
MACE-RI appears to show the most detail with transverse cracks sharpened.
The object appears to be delaminating under impact. 
Cross referencing the cracks with the view data confirms that these cracks are real
and not residual streak artifacts.

    
\section{CONCLUSION}
\label{sec:conclusion}
In this paper, we demonstrated the versatility of MACE 
by tailoring an advanced 3D prior model to address the specific needs of 
the sparse view flash X-ray CT system.
We envision a tunable system where end users can combine different agents as modules 
into the MACE framework to build expressive priors
that can exploit geometries of specimens beyond rotational invariance.






\bibliographystyle{IEEEbib}
\bibliography{refs}

\end{document}